\documentclass[conference,final,10pt,letterpaper]{IEEEtran}
\usepackage{amsmath,amssymb,amsfonts}
\usepackage{algorithmic}
\usepackage{subfig}
\usepackage{graphicx}
\usepackage{textcomp}
\usepackage{xcolor}
\usepackage[numbers]{natbib}
\usepackage{url}
\usepackage{aas_macros}  
\usepackage{tikz}
\usepackage{xspace}
\usepackage{multirow}
\usepackage{verbatim}  
\usepackage{placeins}
\usepackage{hyperref}

\DeclareRobustCommand{\IEEEauthorrefmark}[1]{\smash{\textsuperscript{\footnotesize #1}}}
\newcommand{\ccglib}{\texttt{ccglib}\xspace}
\newcommand{\cudapeak}{\texttt{cudapeak}\xspace}
\begin{document}
\bstctlcite{BSTcontrol}

\title{The Tensor-Core Beamformer: A High-Speed Signal-Processing Library for Multidisciplinary Use\\
}

\author{
    \IEEEauthorblockN{Leon Oostrum\IEEEauthorrefmark{1}, Bram Veenboer\IEEEauthorrefmark{2}, Ronald Rook\IEEEauthorrefmark{3}, Michael Brown\IEEEauthorrefmark{4}, Pieter Kruizinga\IEEEauthorrefmark{4}, John W. Romein\IEEEauthorrefmark{2}}
    \IEEEauthorblockA{\IEEEauthorrefmark{1}Netherlands eScience Center, Amsterdam, the Netherlands, l.oostrum@esciencecenter.nl}
    \IEEEauthorblockA{\IEEEauthorrefmark{2}ASTRON (Netherlands Institute for Radio Astronomy), Dwingeloo, the Netherlands,
    \{veenboer, romein\}@astron.nl}
    \IEEEauthorblockA{\IEEEauthorrefmark{3}Sioux Technologies, Eindhoven, the Netherlands}
    \IEEEauthorblockA{\IEEEauthorrefmark{4}Erasmus Medical Center, Rotterdam, the Netherlands}
}
\maketitle

\begin{abstract}
Beamforming is a well-known technique to combine signals from multiple sensors. It has a wide range of application domains.
This paper introduces the Tensor-Core Beamformer: a generic, optimized beamformer library that harnesses the computational power of GPU tensor cores to accelerate beamforming computations.
The library hides the complexity of tensor cores from the user, and supports 16-bit and 1-bit precision.
An extensive performance evaluation on NVIDIA and AMD GPUs shows that the library outperforms traditional beamforming on regular GPU cores by a wide margin, at much higher energy efficiency.
In the 16-bit mode, it achieves over 600 TeraOps/s on an AMD MI300X GPU, while approaching 1 TeraOp/J. In the 1-bit mode, it breaks the 3 PetaOps/s barrier and achieves over 10 TeraOps/J on an NVIDIA A100 GPU.
The beamforming library can be easily integrated into existing pipelines. We demonstrate its use for medical ultrasound and radio-astronomical instruments.
\end{abstract}


\section{Introduction}
\label{sec:intro}
Beamforming is one of the basic ways to combine the signals from multiple sensors.
It is a well-known technique to increase sensitivity in specified directions (the beams), and suppress signals from other directions.
Beamforming can be used for sound and radio waves, for transmissions and reception, and can be performed on analog and digitized signals.
The applications range from wireless communication to seismology.

This paper focuses mostly on digital beamforming.
Depending on the application, the data rate of signals, the number of sensors, or the number of beams can be large.
This leads to a compute challenge, while the signals typically need to be processed in real time, and often under constraints that limit the energy use.

In this paper, we explore the use of \emph{tensor cores} for beamforming. Tensor cores are hardware matrix-matrix multiplication units found in most modern Graphics Processing Units (GPUs).
They perform these multiplications much more efficiently than regular GPU cores, but typically only work for limited-precision input data (e.g. 8-bit integers), while the output is usually 32-bit integer or floating point.
Tensor cores are a key technology for accelerating training and inference algorithms in deep learning.
However, they can be used for any algorithm that can be expressed as matrix-matrix multiplications and operates on limited-precision input data.
Beamforming is one of them, provided that multiple beams (i.e. directions or positions) are created from the same input streams, and that the weights used to steer the beams are constant for some period of time (they should not be different for every sample in time).
Under these conditions, the algorithm is a multiplication of a matrix of sampled data by a matrix of weights (see also Section~\ref{sec:background}).
The input data normally come from Analog-to-Digital Converters of which the accuracy is limited to a dozen bits or less, hence there is no benefit from performing the beamforming multiplications in high precision.

The main contributions of this paper are the introduction and performance analysis of the \emph{Tensor-Core Beamformer} (TCBF),
a highly efficient beamforming library for multiple application domains.
This library is domain independent, and performs complex-valued matrix-matrix multiplications, hiding the complexity of using tensor cores.
In addition, we integrated the library into beamforming applications for radio-astronomical use and medical computational ultrasound imaging use. In these applications, the TCBF is up to a factor 10-100 faster than previous GPU-based beamforming implementations, as well as an order of magnitude more energy efficient.
This paper describes and evaluates the library, as well as the use in both application domains.

The paper is structured as follows:
Section~\ref{sec:background} provides an overview of the background and related work. In Section~\ref{sec:tcbf}, we introduce the TCBF and discuss its key features. Section~\ref{sec:performance} delves into auto-tuning techniques for optimizing performance and energy efficiency, followed by analysis of the optimized TCBF. Next, the applications of the TCBF in radio astronomy and medical ultrasound are detailed in Section~\ref{sec:applications}. Finally, we conclude with a summary of findings and future directions in Section~\ref{sec:conclusion}.

\section{Background and related work}
\label{sec:background}
Beamforming is a signal-processing technique used to direct the reception or transmission of signals in a specific direction by combining signals from multiple sensors, typically in an array configuration~\citep{Veen1988}. This process enhances the signal-to-noise ratio (SNR) for the signal of interest while suppressing interference from other directions.
Beamforming is for example used to search for pulsars or Fast Radio Bursts~\citep{leeuwen2023} in radio astronomy, and for medical ultrasound imaging~\citep{Perrot2021}.

Consider an array of $N$ sensors receiving a plane wave signal $s(t)$ from a far-field source at an angle $\theta$. The signal received by the $k$-th sensor can be expressed as:
\begin{equation}
x_k(t) = s(t - \tau_k) + \sigma_k(t)    
\end{equation}

where $\tau_k$ is the time delay associated with the $k$-th sensor due to the wavefront's angle of arrival, and $\sigma_k(t)$ is the noise at the $k$-th sensor. The time delay $\tau_k$ is given by:
\begin{equation}
\tau_k = \frac{d_k \sin\theta}{c}
\end{equation}

where $d_k$ is the distance of the $k$-th sensor from a reference point in the array, and $c$ is the speed of the wave (e.g. speed of light for electromagnetic waves, speed of sound for acoustic waves).

In beamforming, the signals received by each sensor are combined with appropriate weights to form the beamformed output $y(t)$:
\begin{equation}
\label{eq:bf}
y(t) = \sum_{k=1}^{K} w_k x_k(t)
\end{equation}

where $w_k$ are the complex weights applied to each sensor's signal. These weights are designed to steer the beam in the desired direction, effectively aligning the phases of the signals from the desired source and canceling out interference from other directions. When multiple samples are beamformed at once, Eq.~\ref{eq:bf} maps to a matrix-matrix multiplication. Matrix-matrix multiplication is typically described as the product of an $M\!\times\!K$ matrix with an $K\!\times\!N$ matrix. The result is an $M\!\times\!N$ matrix. In the beamforming algorithm, M corresponds to the number of beams, N is the number of samples beamformed at a time, and K is the number of elements that is summed over, i.e. the number of receivers in the above description.

The beamforming procedure becomes more complicated for near-field sources and when transmissions of an (acoustic) signal are included as well, but the overall procedure remains the same and can still be mapped to a matrix-matrix multiplication.

Tensor cores have been used before for signal processing: The Tensor-Core Correlator~\citep{Romein:21a} is a highly (energy) efficient library that correlates the signals from multiple radio telescopes.
Correlating is the ``other'' method to combine the signals from multiple sensors, and is, for example, used to create sky images. Whereas beamforming is a weighted addition of all sensor signals, correlations are pair-wise multiplications.

Highly-efficient GPU matrix-matrix multiplication libraries already exist, such as CUTLASS and cuBLAS/rocBLAS. However, these libraries typically impose restrictions on data layouts and some lack full support for 1-bit and complex-valued computations.
The lower-level Warp Matrix Multiply-Accumulate interface allows precise control over data placement in shared and device memory, enabling a custom kernel optimized for our needs.

Recent AMD GPUs have their own version of tensor cores, called matrix cores. The domain-independent layer of the tensor core beamformer supports AMD matrix cores as well. Although this paper typically uses NVIDIA terminology, the content applies to the AMD equivalent as well. The only exception is 1-bit precision, which is only supported on NVIDIA GPUs.

\section{The Tensor-Core Beamformer}
\label{sec:tcbf}
The core of the beamforming algorithm is a complex-valued matrix-matrix multiplication, which we have implemented in a separate library, \ccglib\footnote{\url{https://git.astron.nl/RD/recruit/ccglib}}.
\ccglib supports both CUDA and HIP. To program the tensor cores of NVIDIA GPUs, we use the Warp Matrix Multiply Accumulate (WMMA) interface. AMD implements a similar interface through rocWMMA, available in HIP. The user of \ccglib can switch between the CUDA and HIP backends with a CMake flag, or when compiling manually simply by switching between the \emph{nvcc} and \emph{hipcc} compilers. The use of the tensor cores and the complexity of supporting both HIP and CUDA is hidden from the user. The user only has to provide the input and output matrices and tell \ccglib what shapes and types the matrices have. To achieve optimal performance, \ccglib compiles the GPU kernel at runtime with knowledge of both the type of GPU used, and of all input parameters such as the number of receivers and the number of beams to be created. It is also possible to execute several matrix-matrix multiplications at once through a batch size option.

\ccglib currently supports 16-bit float and 1-bit integer precision. We focus on 16-bit float processing because some domain-specific input data is naturally in this format, making it both practical and efficient. 16-bit tensor-core operations offer significant speedups over 32-bit float on standard GPU cores, while also halving memory usage and bandwidth requirements.

1-bit processing enables higher throughput by reducing memory bandwidth and increasing arithmetic intensity, as the same number of operations are performed on fewer bits. Since 1-bit arithmetic is faster than 16-bit, this can lead to significant speedups. While lower precision introduces quantization noise, beamforming remains robust since many values are accumulated. By leveraging tensor cores for 1-bit arithmetic, we explore its potential for efficient, high-performance beamforming.

In addition to a matrix-matrix multiplication GPU kernel, \ccglib implements two more types of kernels: For 1-bit precision, the input data must be packed, i.e. 32 consecutive 1-bit samples must be stored in a single 32-bit integer. Packing and unpacking kernels are provided to handle this. Additionally, the matrix-matrix multiplication kernel requires that the input matrices are tiled in device memory. This can be handled by \ccglib through a transpose kernel. The packing and transpose kernels are relatively straightforward, and both are bound by memory bandwidth as they only move data around.

Beamforming for a specific scientific domain can be implemented as a thin wrapper around \ccglib. Two such applications are described in Sect.~\ref{sec:applications}. 

During the implementation of \ccglib, we identified several challenges: the absence of support for complex numbers, the details of 1-bit arithmetic, the limited support for 1-bit tensor-core operations by the NVIDIA Hopper GPU architecture, and the fact that tensor cores have such a high compute throughput that it is difficult to feed them data fast enough. Before discussing these challenges, we investigate the potential of tensor-core technology through a set of micro-benchmarks on a range of workstation and server-grade GPUs: NVIDIA's RTX 4000 Ada (Hereafter AD4000), Tesla A100 (A100), and Grace Hopper (GH200), as well as AMD's Radeon Pro W7700 (W7700), Instinct MI210 (MI210), Instinct MI300X (MI300X), and Instinct MI300A (MI300A).

\subsection{Tensor-core micro-benchmarks}
\label{sec:bench}
Tensor cores support several precisions and matrix sizes, and different GPU architectures support different combinations of these parameters. To get an overview of the attainable tensor core performance, we have run micro-benchmarks on several GPU architectures. These micro-benchmarks do not load data from global memory, to avoid memory throughput bottlenecks (see also Sect.~\ref{sec:data_reuse}). The benchmarks were run using the \cudapeak\footnote{https://gitlab.com/astron-misc/cudapeak/} library. The results are summarized in Table~\ref{tab:bench}. 

When computing the peak performance using the measured clock frequency, which differs from the theoretical maximum, \cudapeak's performance is close to the peak on all GPUs, except for the GH200, which falls notably short. The GH200 and other GPUs of the Hopper generation support a new interface to the tensor cores, called WGMMA. Only with this interface, it is possible to reach maximum performance. As shown in~\citep{luo2024}, the WMMA interface limits the performance to $60-65\%$ of the maximum. Our benchmark indeed reaches $\sim65\%$ of the expected GH200 peak performance.

For 1-bit precision, the $16\!\times\!8\!\times\!256$ matrix fragment layout is not available through the WMMA interface, only through inline PTX. In both \cudapeak and \ccglib, we have included an extension to WMMA with support for this fragment layout. The 1-bit benchmarks were run with both the WMMA-supported layout of $8\!\times\!8\!\times\!128$ as well as with this custom extension, and with both XOR and AND as multiplication operand. This leads to a total of four different benchmarks. These are only run on NVIDIA GPUs, as 1-bit matrix values are not supported on AMD GPUs. 

The $8\!\times\!8\!\times\!128$ and $16\!\times\!8\!\times\!256$ layouts have the same performance on the AD4000, but on the A100 and GH200 the larger layout is at least twice as fast. As the larger layout is never slower than the smaller one, there seems to be no reason to use the small layout when considering just the tensor core throughput.
We also note that on the GH200, using XOR as an operand is up to five times slower than using AND. The CUDA documentation notes that XOR is deprecated as of the Hopper generation. However, the instruction is still available at both the WMMA and PTX level. Inspecting the generated SASS assembly reveals that the XOR operation has been removed from hardware, and in software it is replaced by several AND operations combined with boolean logic. This is the reason for the low performance of the XOR mode on the GH200. In the best-performing case, we see the same $\sim65\%$ of maximum performance as for float16, resulting from our use of the WMMA interface instead of WGMMA.

\begin{table*}[tbp]
\centering
\caption{Tensor core micro-benchmark results for 16-bit float and 1-bit integer precision. The measured tensor core throughput as well as the theoretical value are shown. 1-bit precision was benchmarked with two matrix fragment layouts and two operands for the multiplication operation. It is available on NVIDIA GPUs only.}
\begin{tabular}{ll|lllllll}
\hline
Input / output type & Fragment size &  \multicolumn{5}{c}{Measured performance / Theoretical peak (TOPs/s)} \\
& $M\!\times\!N\!\times\!K$& \multicolumn{1}{c}{AD4000$^a$} & \multicolumn{1}{c}{A100} & \multicolumn{1}{c}{GH200} & \multicolumn{1}{c}{W7700$^a$} & \multicolumn{1}{c}{MI210} & \multicolumn{1}{c}{MI300X$^b$} & \multicolumn{1}{c}{MI300A$^b$} \\
\hline
float16 / float32  & $16\times16\times\phantom{0}16$ & \phantom{0}117 / \phantom{0}107 & \phantom{0}308 / \phantom{0}312  & \phantom{00}646 / \phantom{00}990 & 59 / 57 & 174 / 181 & 1205 / 1307 & 949 / 981 \\
int1 / int32 (XOR) & $\phantom{0}8\times\phantom{0}8\times128$  & 1847 / 1710 & 2465  / 4992 & \phantom{00}979 / 15800$^c$ & N/A & N/A & N/A & N/A \\
int1 / int32 (AND) & $\phantom{0}8\times\phantom{0}8\times128$  & 1804 / 1710 & 2408  / 4992 & \phantom{0}3894 / 15800$^c$ & N/A & N/A & N/A & N/A \\
int1 / int32 (XOR) & $16\times\phantom{0}8\times256$ & 1865 / 1710 & 4942  / 4992 & \phantom{0}2361 / 15800$^c$ & N/A & N/A & N/A & N/A \\
int1 / int32 (AND) & $16\times\phantom{0}8\times256$ & 1865 / 1710 & 4942  / 4992 & 10276 / 15800$^c$ & N/A & N/A & N/A & N/A \\
\end{tabular}
\label{tab:bench}
\\\vspace{4pt}\footnotesize{$^a$The AD4000 and W7700 run at boosted clock speeds beyond their vendor's specification, explaining why they perform better than the theoretical maximum. $^b$The MI300X and MI300A cannot sustain the maximum clock speed in this synthetic benchmark, leading to lower performance than the theoretical value.
$^c$NVIDIA does not provide the theoretical 1-bit performance for the GH200. We assume it scales from float16 the same as on the Ampere and Ada generation GPUs.}
\end{table*}

\subsection{Complex number support}
Tensor cores were created to accelerate common computations in deep learning and are only capable of executing real-valued matrix-matrix multiplications. Additionally, they only provide an accumulation operation, subtraction is not available. To implement the multiplication of two complex numbers on tensor cores, we need both, though. 

Given two complex numbers $a$ and $b$, complex multiplication is defined as follows:
$$\operatorname{Re}(a \times b)= \operatorname{Re}(a) \operatorname{Re}(b) - \operatorname{Im}(a) \operatorname{Im}(b)$$
$$\operatorname{Im}(a \times b)= \operatorname{Re}(a) \operatorname{Im}(b) + \operatorname{Im}(a) \operatorname{Re}(b)$$

Starting with output initialized to zero, this can be implemented for matrices on the tensor cores in five steps:
\begin{enumerate}
    \item $\operatorname{Re}(a \times b) \mathrel{+}= \operatorname{Re}(a) \operatorname{Re}(b)$ 
    \item $\operatorname{Im}(a \times b) \mathrel{+}= \operatorname{Re}(a) \operatorname{Im}(b)$ 
    \item $\operatorname{Im}(b)\phantom{a\times +} = -\operatorname{Im}(b)$
    \item $\operatorname{Re}(a \times b) \mathrel{+}= \operatorname{Im}(a) \operatorname{Im}(b)$ 
    \item $\operatorname{Im}(a \times b) \mathrel{+}= \operatorname{Im}(a) \operatorname{Re}(b)$ 
\end{enumerate}

Hence, complex matrix-matrix multiplication can be implemented using four real-valued matrix-matrix multiplications and one negation of the imaginary part of the $b$ matrix. The negation of $\operatorname{Im}(b)$ is executed in local registers, so it is fast and does not modify the global input data. 

\subsection{Need for data reuse}
\label{sec:data_reuse}
To achieve good performance on tensor cores, it is of utmost importance to ensure the data are efficiently reused throughout the GPU memory hierarchy, from global memory, L2 and L1 caches, and shared memory, to registers.

The GPU kernels in \ccglib are adaptive in the amount of work per thread block and warp, which affects the amount of reuse at the shared-memory level and register-file level, respectively. Optimal configurations for specific GPUs have been determined through auto-tuning as described in Sect.~\ref{sec:tuning}. \ccglib automatically selects these parameters at runtime, the user does not need to provide them. Through this mechanism, we ensure optimal reuse of data at all levels of memory.

In addition to data reuse, we can reduce memory bottlenecks by making use of asynchronous data copies between GPU global and shared memory, available on NVIDIA Ampere and later GPUs. With this feature, it is possible to overlap computations with data transfer to reduce execution time. We implement this by creating a multi-stage buffer in shared memory. While data is being copied to one buffer, another buffer can be copied to the register file and used for computations. Using the CUDA pipeline synchronization primitives, we ensure that data has been written to a shared memory buffer before it is read by the threads and written to the register file. The number of buffers is tunable and automatically set to one on AMD GPUs, which do not support these asynchronous copies.

\subsection{1-bit arithmetic on tensor cores}
In a 1-bit representation of a real number, only two possible values exist. A natural choice is to use these two values to represent $-1$ and $1$, as they preserve sign information. Importantly, this implies that the number $0$ cannot be represented.

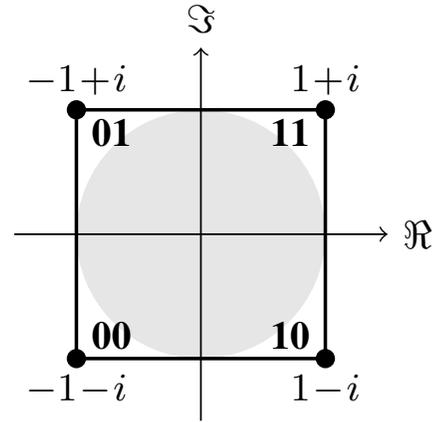
\begin{figure}
\resizebox{0.7\columnwidth}{!}{
\begin{tikzpicture}
    \draw[draw=none, fill=black, dashed, opacity=0.1] (0,0) circle(1);
    
    \draw[thick] (-1,-1) rectangle (1,1);
    
    \draw[->] (-1.5,0) -- (1.5,0) node[right] {\small $\Re$};
    \draw[->] (0,-1.5) -- (0,1.5) node[above] {\small $\Im$};

    \filldraw (1,1) circle(2pt);  
    \filldraw (-1,1) circle(2pt); 
    \filldraw (-1,-1) circle(2pt); 
    \filldraw (1,-1) circle(2pt);  

    \node[anchor=south] at (1, 1) {\small $1\!+\!i$}; 
    \node[anchor=south] at (-1, 1) {\small $-1\!+\!i$};
    \node[anchor=north] at (-1, -1) {\small $-1\!-\!i$};
    \node[anchor=north] at (1, -1) {\small $1\!-\!i$};

    \node[anchor=north west] at (-1, 1.05) {\small \texttt{\bf 01}};
    \node[anchor=north east] at (1, 1.05) {\small \texttt{\bf 11}};
    \node[anchor=south west] at (-1, -1.05) {\small \texttt{\bf 00}};
    \node[anchor=south east] at (1, -1.05) {\small \texttt{\bf 10}};
\end{tikzpicture}
}
\centering
\caption{1-bit complex numbers and their binary representation. The representable values 
$-1\!-\!i$, $-1\!+\!i$, $1\!-\!i$, and $1\!+\!i$ are shown at the corners of the square, with binary values \texttt{00}, \texttt{01}, \texttt{10}, and \texttt{11}, respectively. The light gray circle represents the unit circle. Note that zero, i.e. $0\!+\!0i$, is not representable.
}
\label{fig:1bit}
\end{figure}

For 1-bit complex numbers, there is one bit per component, meaning that both the real and imaginary parts are independently represented using a single bit. Preserving sign information along both the real and imaginary axes, this results in four possible complex values. These values are equally spread around the unit circle in the complex plane, as shown in Fig.~\ref{fig:1bit}.

NVIDIA tensor cores support 1-bit precision matrix-matrix multiplication using binary operations. Instead of numerical multiplication, they perform a bitwise operation between two input matrices, followed by a population count ($\mathrm{popc}$), which counts the number of bits set to one in the result. The bitwise operation is either XOR (deprecated as of the Hopper architecture) or AND (introduced with the Ampere architecture).

Real-valued matrix-matrix multiplication with the encoding described above can be implemented efficiently using XOR as the bitwise operation. To illustrate this, consider the dot product of two vectors $A$ and $B$ of length $K$, as shown in Table~\ref{tab:1bit_gemm_real} for $K\!=\!4$. The left half of the table shows the vector dot product in decimal. After performing element-wise multiplication, the final value of the dot product is obtained by summing the results.

In the binary case, shown in the right half of the table, the process differs slightly. Instead of multiplying the decimal values directly, we perform an element-wise XOR operation between the corresponding elements of the two vectors. This produces a new vector where a binary zero represents a positive value and a binary one represents a negative value. Due to the XOR operation, the binary encoding in this resulting vector is flipped with respect to the binary encoding in the input vector. The final value of the dot product is then determined by subtracting the number of binary zeroes from the number of binary ones in the resulting vector. The number of ones is counted using the population count $\mathrm{popc}$ function, while the number of zeroes is simply $K$ minus the number of ones.
Thus, the final expression for the 1-bit vector dot product is:
\begin{equation}
K - 2 \, \mathrm{popc}(A \oplus B),
\end{equation}
where $\oplus$ denotes the element-wise XOR operation. Since matrix-matrix multiplication can be expressed as a series of vector dot products, this approach can also be applied to implement matrix-matrix multiplication with 1-bit precision.

\begin{table}[htbp]
\centering
\caption{Vector dot product in 1-bit precision. Every row in this table corresponds to one element of the input vector, i.e. the input A of length $4$ ($K\!=\!4$) has decimal values $1$, $-1$, $1$ and $-1$ and is represented as 1010 in binary.}
\begin{tabular}{ccc|ccc}
\hline
\multicolumn{3}{c|}{Decimal} & \multicolumn{3}{c}{Binary} \\
$A$ & $B$ & $A_k \times B_k$ & $A$ & $B$ & $A_k \oplus B_k$ \\
\hline
\phantom{$-$}1 & \phantom{$-$}1 & \phantom{$-$}1 & 1 & 1 & 0 \\
$-$1 & \phantom{$-$}1 & $-$1 & 0 & 1 & 1 \\
\phantom{$-$}1 & $-$1 & $-$1 & 1 & 0 & 1 \\
$-$1 & $-$1 & \phantom{$-$}1 & 0 & 0 & 0 \\
\hline
\multicolumn{2}{c}{\multirow{2}{*}{$\sum{A_k \times B_k}$}} & \multirow{2}{*}{\phantom{$-$}0} &
\multicolumn{2}{c}{$\phantom{K - 2 }\mathrm{popc}(A \oplus B$)} & 2 \\
& & & \multicolumn{2}{c}{$K - 2 \mathrm{popc}(A \oplus B)$} & 0 \\
\end{tabular}
\label{tab:1bit_gemm_real}
\end{table}

For complex-valued matrix-matrix multiplication, we need to consider two things: the number of matrix multiplications executed and the absence of a representation of zero in the input values.

Firstly, to compute the real and imaginary parts of the output, two separate real-valued matrix-matrix multiplications are required. This means that instead of $K$ terms, $2K$ terms are summed for each part. 

Secondly, when multiplying matrices that do not exactly match the sizes supported by the tensor cores, padding is applied to make the matrix dimensions compatible.
The padded area is typically set to zero. However, zero cannot be represented in 1-bit mode. Instead, we set the padded region to binary $0$, which corresponds to decimal $-1$. This introduces an additional effect that must be accounted for. For the real part of the output, the results of the two matrix-matrix multiplications cancel out the padding effect, as the results are subtracted from each other. For the imaginary part, however, the padding effect leads to an erroneous addition of $K_\mathrm{pad}\!\times\!-1\!\times\!-1$ in the result, where $K_\mathrm{pad}$ denotes the amount of padding. Combining these two effects, we arrive at the following two equations for complex-valued 1-bit matrix-matrix multiplication on tensor cores, using subscript $\mathrm{r}$ and $\mathrm{i}$ to denote the real and imaginary parts:

\vspace{-10pt}
\begin{equation}
(A \times B)_\mathrm{r} = 2\Bigl(K - \bigl(\mathrm{popc}(A_\mathrm{r} \oplus B_\mathrm{r}) + \mathrm{popc}(A_\mathrm{i} \oplus \overline{B_\mathrm{i}})\bigr)\Bigr)
\end{equation}
\begin{equation*}
(A \times B)_\mathrm{i} = 2\Bigl(K - K_\mathrm{pad} - \bigl(\mathrm{popc}(A_\mathrm{r} \oplus B_\mathrm{i})
+\mathrm{popc}(A_\mathrm{i} \oplus 
B_\mathrm{r})\bigr)\Bigr)
\end{equation*}

\subsection{NVIDIA Hopper support}
\label{sec:hopper}
As explained in the previous section, 1-bit matrix-matrix multiplication can be efficiently implemented with an XOR operation. However, this operation is deprecated as of the Hopper architecture and leads to low performance, as discussed in Sect.~\ref{sec:bench} and shown in Table~\ref{tab:bench}. 

To optimize performance on Hopper, we switch to using the AND operation. The XOR operation detects when two input bits are different: if they are different, the output is set to 1; otherwise, it is set to 0. We can achieve similar functionality with the AND operation by following a different sequence of steps. Specifically, we perform an AND operation on the inputs, followed by negating both inputs, then performing another AND operation, and finally summing the results of both AND operations. This method detects when the input bits are the same, as opposed to when they are different. 

This means the (signed) output of the matrix-matrix multiplication is negated relative to the XOR version. In summary, (real-valued) 1-bit matrix-matrix multiplication can be implemented with the AND instruction as follows:
\begin{equation}
2 \bigl(\mathrm{popc}(A\,\land\,B) + \mathrm{popc}(\overline{A}\,\land\,\overline{B})\bigr) - K,
\end{equation}
where $\land$ denotes the element-wise AND operation.
Although using the AND operation requires twice as many tensor core instructions compared to XOR, this still results in a net performance improvement on Hopper because the AND operation is up to five times faster than XOR on this architecture. Therefore, \ccglib automatically switches to the AND-based matrix-matrix multiplication when a Hopper or newer NVIDIA GPU is detected.

\section{Performance and energy efficiency}
\label{sec:performance}
\subsection{Auto-tuning}
\label{sec:tuning}
GPU kernels can typically be run on a GPU with many different thread block dimensions, that all give the correct result but can result in vastly different performance. Additionally, the GPU kernels in \ccglib were designed such that parameters, like the amount of work per thread block and warp, can be set at compile time. Because \ccglib compiles the GPU kernels when the application is running on the host, it can pick different values for these tunable parameters based on the type of GPU used as well as the input data sizes. To find the optimum of the tunable parameters, we need to explore a vast search space, and this process has to be repeated for each GPU architecture. To facilitate this, we use Kernel Tuner~\citep{kerneltuner}, a Python-based auto-tuning framework that can automatically optimize kernels written in both CUDA and HIP~\citep{kerneltunerhip}. Kernel Tuner measures the run time of each configuration of a GPU kernel. It is possible to extend Kernel Tuner with other metrics, either built-in or custom. In addition to performance metrics, we measure the energy consumption of the GPU using the \texttt{Power Measurement Toolkit (PMT)}~\citep{corda2022}. \texttt{PMT} supports power measurements of both NVIDIA GPUs through \texttt{NVML}, as well as AMD GPUs through \texttt{rocm-smi}. 

In \ccglib we have three types of GPU kernels: a packing kernel for 1-bit data, a transpose kernel, and matrix-matrix-multiplication kernels for 16-bit float and 1-bit integer types. Only the matrix-matrix multiplication kernel is always invoked. The use of the others depends on the earlier steps of the processing pipeline \ccglib is used in. Additionally, the matrix-matrix multiplication kernels take most time and are the only kernels that have tunable parameters other than the thread block size. Hence, we focus our optimization process on these kernels.

The optimal tuning parameters do not only depend on the GPU, but also on the size of the input and output data as well as the precision used. As a generic use case, we tune the float16 kernel for $M\!=\!N\!=\!K\!=\!8192$, while for 1-bit integer we select $M\!=\!32768$, $N\!=\!8192$, and $K\!=\!524288$. To assess a kernel, we define the performance in TOPs/s as the number of useful operations, i.e. $8\!\times\!M\!\times\!N\!\times\!K$, per second. In the limit of large matrices, the product of the matrix sizes is the number of fused multiply-add (FMA) instructions required for real-valued matrix-matrix multiplication. The factor eights comes from the fact that four FMA instructions are required for each complex multiplication, and each FMA counts as two instructions. The resulting performance number is divided by the average power consumption of the GPU during the kernel execution to obtain the number of operations per second per Watt, or equivalently the number of operations per Joule.

\begin{figure*}[tbp]
\includegraphics[width=\textwidth]{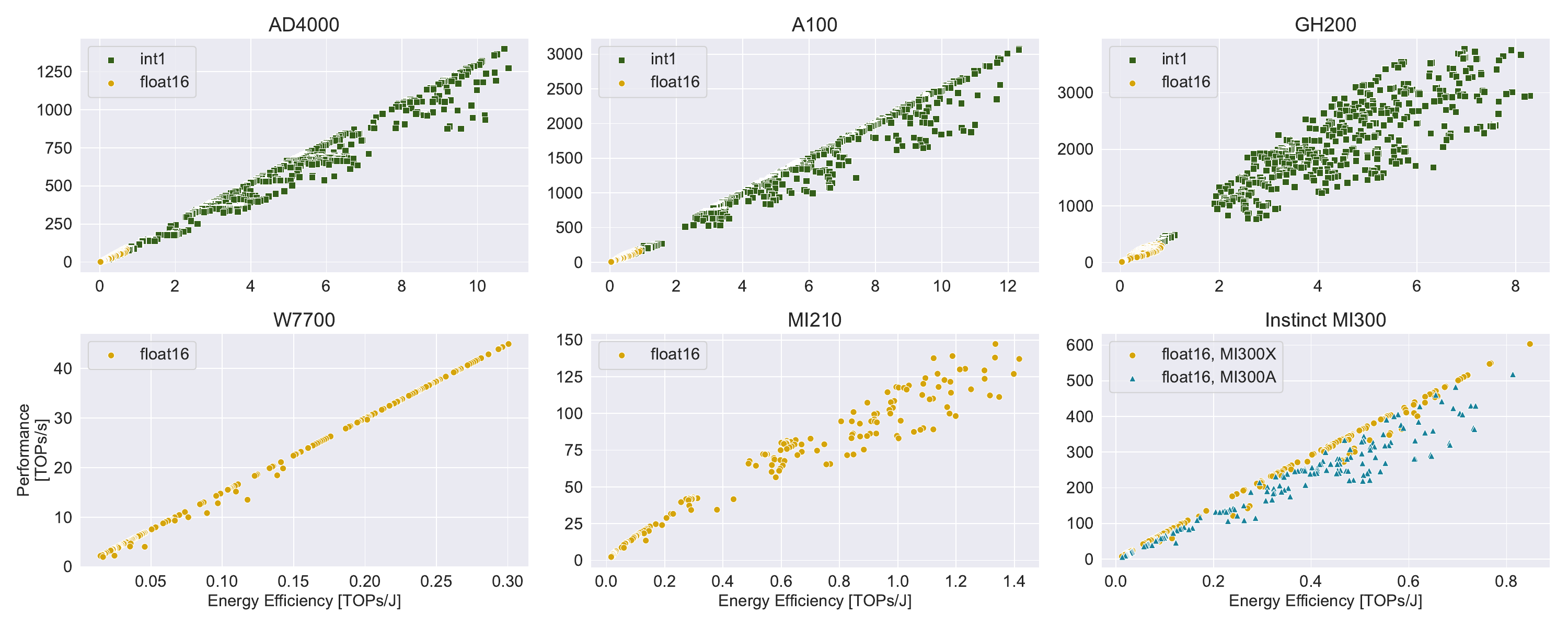}
\caption{Auto-tuning results of \ccglib matrix-matrix multiplication kernel. The measured performance and energy efficiency of each combination of tuning parameters is shown.}
\label{fig:tuning}
\end{figure*}

The performance and energy efficiency of each combination of tuning parameters is shown in Fig.~\ref{fig:tuning}. Typically, the most performant combination of parameters is also the most energy efficient solution. On the GH200, there is a large spread of kernels with similar performance, but up to a factor two difference in energy efficiency. 

A summary of the parameters for the fastest kernels is given in Table~\ref{tab:tuning}. In float16, the MI300X is both the fastest and most energy-efficient GPU. The GH200 is the fastest in int1, although the A100 is more energy efficient. The optimal tuning parameter values typically vary a lot from GPU to GPU. The MI300X and MI300A optimal parameters are identical, which is not surprising given that they are built using identical architectures but with a different number of accelerator complex dies. While a default set of parameters is shipped with \ccglib, a GPU-specific optimization is best.

\begin{table*}[tbp]
\centering
\caption{Matrix-matrix multiplication kernel performance, energy efficiency, and optimal tuning parameter values.}
\begin{tabular}{lllllllll}
\hline
GPU & Precision & TOPs/s & TOPs/J & M per block & M per warp & N per block & N per warp & Number of buffers \\
\hline
AD4000 & float16 & \phantom{00}93 & \phantom{0}0.7 & 256 & \phantom{0}32 & \phantom{0}32 & 32 & 2\\
A100 & float16 & \phantom{0}173 & \phantom{0}0.8 & 256 & \phantom{0}64 & \phantom{0}32 & 32 & 2\\
GH200 & float16 & \phantom{0}335 & \phantom{0}0.8 & 128 & \phantom{0}64 & \phantom{0}64 & 32 & 2\\
W7700 & float16 & \phantom{00}45 & \phantom{0}0.3 & 256 & 128 & \phantom{0}64 & 16 & 1\\
MI210 & float16 & \phantom{0}147 & \phantom{0}1.3 & 128 & \phantom{0}64 & \phantom{0}64 & 32 & 1\\
MI300X & float16 & \phantom{0}603 & \phantom{0}0.9 & 128 & \phantom{0}64 & 128 & 32 & 1\\
MI300A & float16 & \phantom{0}518 & \phantom{0}0.8 & 128 & \phantom{0}64 & 128 & 32 & 1\\
\hline
AD4000 & int1 & 1400 & 10.7 & 256 & 128 & \phantom{0}32 & 16 & 2\\
A100 & int1 & 3080 & 12.3 & 128 & \phantom{0}32 & \phantom{0}64 & 64 & 4\\
GH200 & int1 & 3780$^a$ & \phantom{0}6.0$^a$ & \phantom{0}64 & \phantom{0}64 & 128 & 32 & 2\\
\hline
\end{tabular}
\\\vspace{4pt}\footnotesize{$^a$ This performance number is with respect to the theoretical amount of useful operations. Because the GH200 uses AND-based tensor-core instructions (see Sect.~\ref{sec:hopper}), which require twice the number of instructions, the actual throughput of the tensor cores is twice as high.}
\label{tab:tuning}
\end{table*}

\subsection{Roofline analysis}
After auto-tuning, we know that we reach the maximum performance obtainable for our implementation of the matrix-matrix multiplication algorithm. However, we also want to assess whether or not we reach good performance relative to each GPU's capabilities. In addition to the maximum tensor core throughput discussed in Sect.~\ref{sec:bench}, we need to consider the memory throughput. This naturally leads to a roofline analysis, where we compare our implementation to the theoretical maximum obtainable on each GPU. To construct the ceiling of the roofline, we use the theoretical memory bandwidth of the GPU and the measured peak tensor core throughput (see Table~\ref{tab:bench}). For both the 16-bit and 1-bit kernels, we then select a small and large matrix size and tune the kernel parameters as described in Sect.~\ref{sec:tuning}, and select the best-performing kernel. The matrix sizes are set as follows ($\mathrm{batch~size}\!\times\!M\!\times\!N\!\times\!K$): float16 small\,-\,$256\!\times\!1024\!\times\!1024\!\times\!64$, float16 big\,-\,$1\!\times\!8192\!\times\!8192\!\times\!8192$, int1 small\,-\,$256\!\times\!1024\!\times\!1024\!\times\!256$, int1 big\,-\,$1\!\times\!32768\!\times\!8192\!\times\!524288$. The performance and number of operations are defined the same way as during the tuning. We then use the theoretical amount of bytes transferred to and from device memory to calculate the arithmetic intensity (AI). 

\begin{figure*}[tbp]
\includegraphics[width=\textwidth]{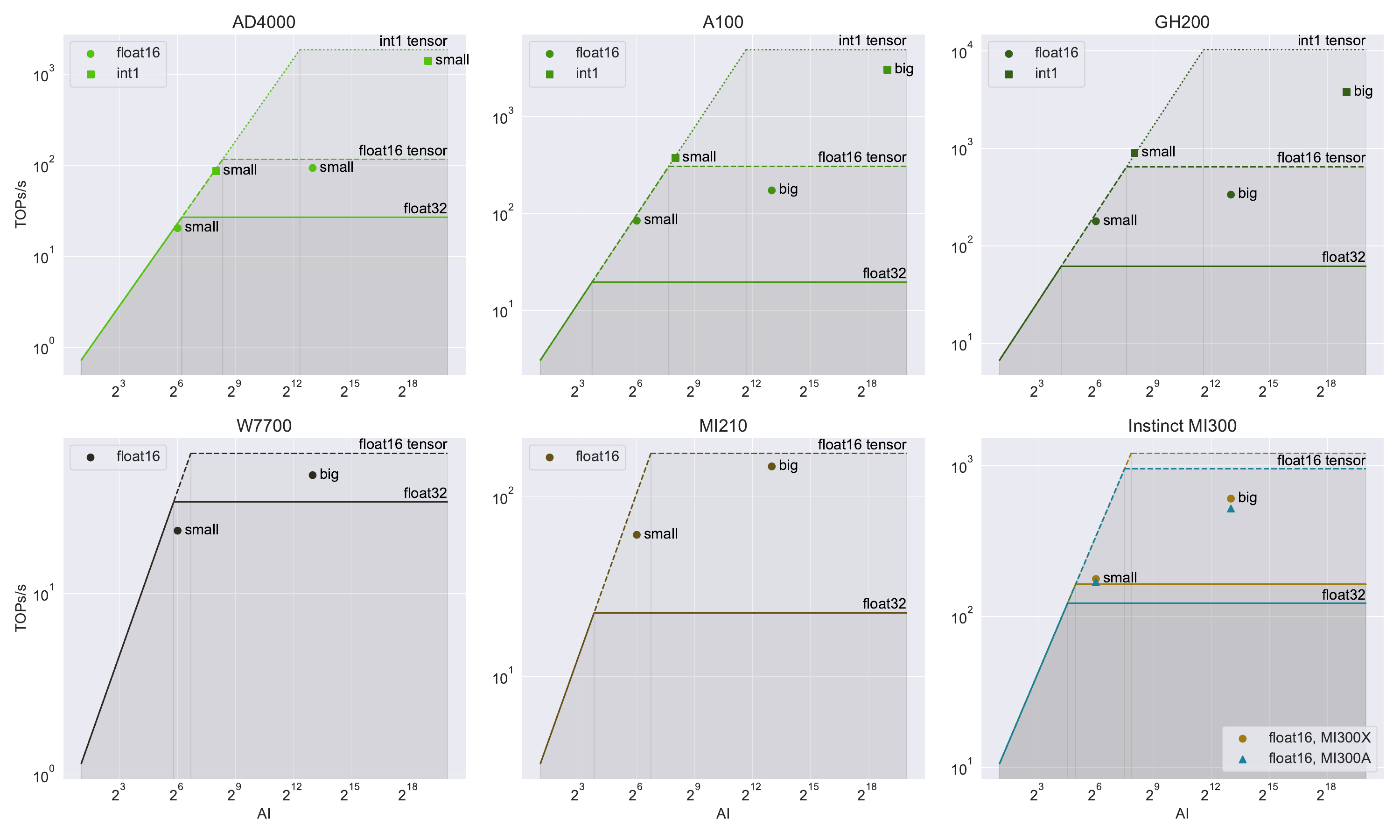}
\caption{Roofline analysis of the \ccglib matrix-matrix multiplication kernel. For each GPU, we show the roofline ceiling of the float16 and int1 (NVIDIA only) tensor cores, as well as the normal float32 cores for comparison.}
\label{fig:roofline}
\end{figure*}

The resulting rooflines are shown in Fig.~\ref{fig:roofline}. For all GPUs, the small matrix size is memory-bound. On all GPUs, but especially the NVIDIA GPUs, we reach a performance very close to the limit set by the memory bandwidth. The larger matrix size is compute bound, and reaches $50-85\%$ of the peak tensor-core throughput.
In all cases except the small matrix size on the workstation-grade GPUs, \ccglib is faster than the theoretical maximum of the normal single-precision cores by a wide margin. 

\subsection{Benchmarking}
We have shown that we reach good performance on a specific set of matrix sizes, however we aim for a solution that is generally applicable and reaches good performance for a wide range of matrix sizes. While it is possible to auto-tune the matrix-matrix multiplication kernel for each potential matrix input size, this is not feasible in practice. Instead, we take the best parameters from Table~\ref{tab:tuning}, and use the built-in benchmark tools of \ccglib to measure performance and energy efficiency across a range of matrix sizes.

\begin{figure*}[tbp]
\centering
\subfloat[16-bit float]{%
    \includegraphics[width=\textwidth]{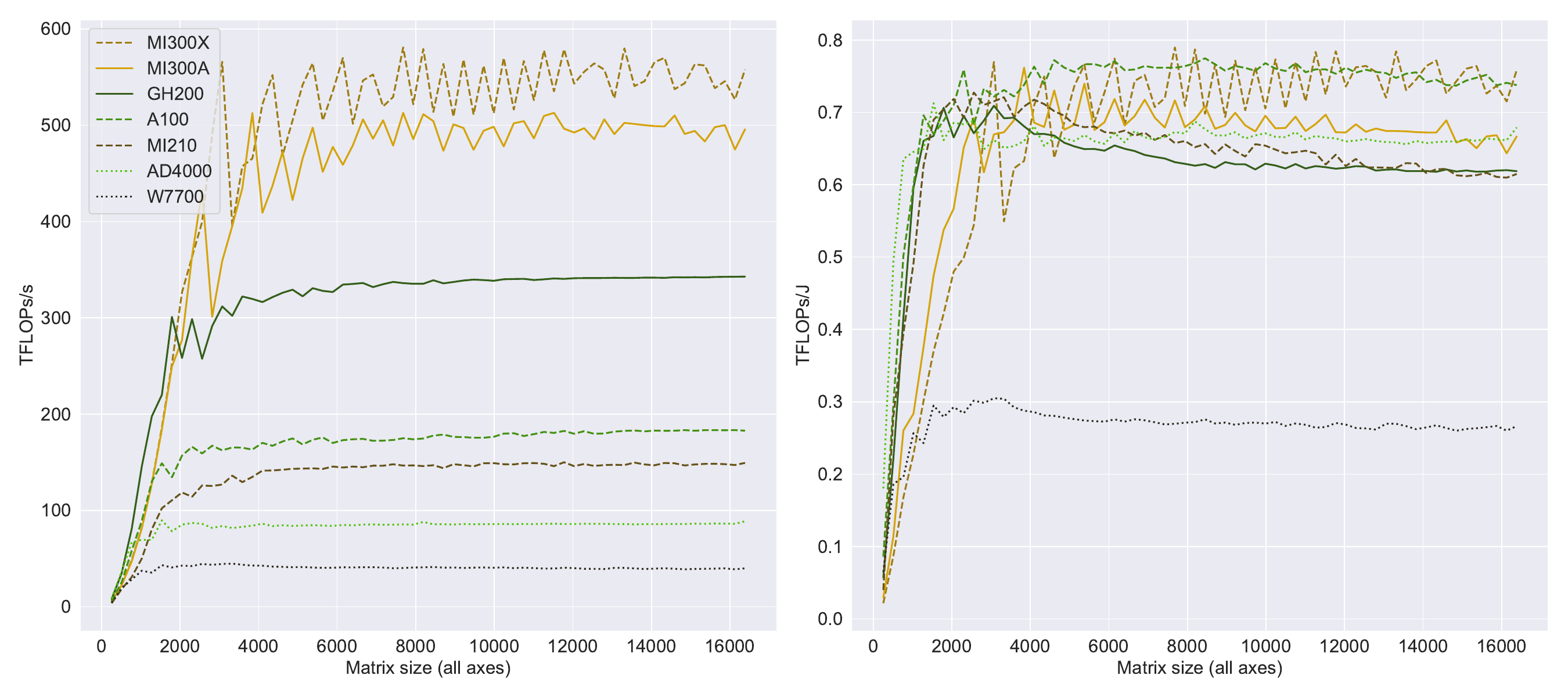}
    \label{fig:benchmark_float16}
}
\newline
\subfloat[1-bit int]{%
    \includegraphics[width=\textwidth]{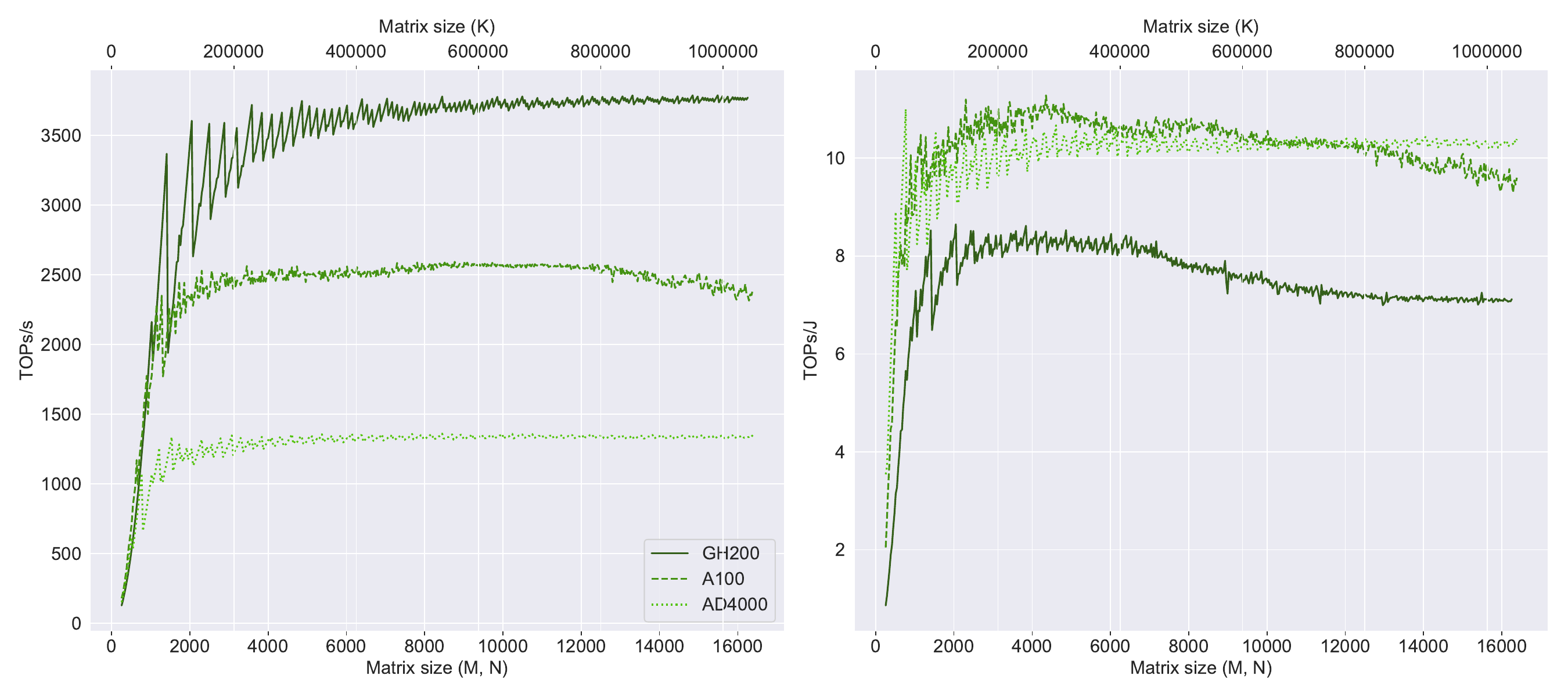}
    \label{fig:benchmark_int1}
}
\caption{Complex matrix-matrix multiplication benchmark results for \protect\subref{fig:benchmark_float16} 16-bit data and \protect\subref{fig:benchmark_int1} 1-bit data. The left panels show performance, while the right panels show energy efficiency.}
\label{fig:benchmark}
\end{figure*}

The results are shown in Fig.~\ref{fig:benchmark}. For all GPUs and precisions, the performance and energy efficiency is substantially lower for smaller matrices. However, starting from matrices of a few thousand elements on each side, we typically reach close to optimal performance. The performance is best when the matrix size is a multiple of the amount of work per thread block. Otherwise, data are padded and the performance is relatively lower. This is the cause of the sawtooth pattern in the results.
Overall, \ccglib performs well on a large range of matrix sizes.

\section{Applications}
\label{sec:applications}
\subsection{Computational ultrasound imaging}
Computational ultrasound imaging (cUSi) is a recent advancement in the field of medical ultrasound. It allows for 3D imaging using a spatially under-sampled transceiver array in conjunction with an spatial encoding mask and a large computational model to decode the spatial information needed to form an image~\citep{KruizingaSciAdv2017}. The cUSi technique essentially changes the sensing problem to a compute problem. This year, Brown et al. showed that using cUSi it is possible to obtain 3D images of blood flow in a mouse brain using an array of only 64 transceivers, which normally requires several thousands of transceivers~\citep{BrownSciAdv2023}. 

However, the caveat of this technique is the number of computations needed to form an image, which makes it currently not possible to obtain real-time imaging feedback. The imaging reconstruction relies on the multiplication of a measurement matrix with an acoustic model matrix which contains for every voxel in the image volume (number of columns) all the expected pulse-echo signals for each transceiver and for each measurement (number of rows). Typically, the minimum number of voxels is $128^3$ and the number of rows for a 64-transceiver probe is 128 (temporal frequencies) $\times$ 64 (transceivers) $\times$ 32 transmissions. The measurement matrix has the same number of rows as the model matrix and the number of columns equals the number of repeated measurements from which, in the case of imaging blood flow, the Doppler signal is computed. This number, which is named ensemble size, can range from 100-10000 frames. In this example case we use $\sim8000$ frames. 

The real-time constraints for this problem are also challenging. Considering a pulse-echo repetition frequency of 32 kHz and an ensemble size of 8000, the time required for the image reconstruction (matrix-matrix multiplication) should be less than 8 seconds in order to maintain real-time feedback. 

In this work we show the use of an ultrasound tensor-core beamformer implemented as a wrapper around \ccglib. We tackle the real-time feedback problem by shrinking the volume size to either a smaller sub-volume, as we do in this example case, or several orthogonal planes through the volume. In addition, we explore a further reduction of the required memory by only keeping the sign of the signal both in the measurement matrix as well as the model matrix. In this approach the data only requires single bit precision. Note that the Doppler processing is done before extracting the sign. Otherwise, the Doppler signal will be lost in the dominant stationary signals.

\begin{figure}[tbp]
\includegraphics[width=\columnwidth]{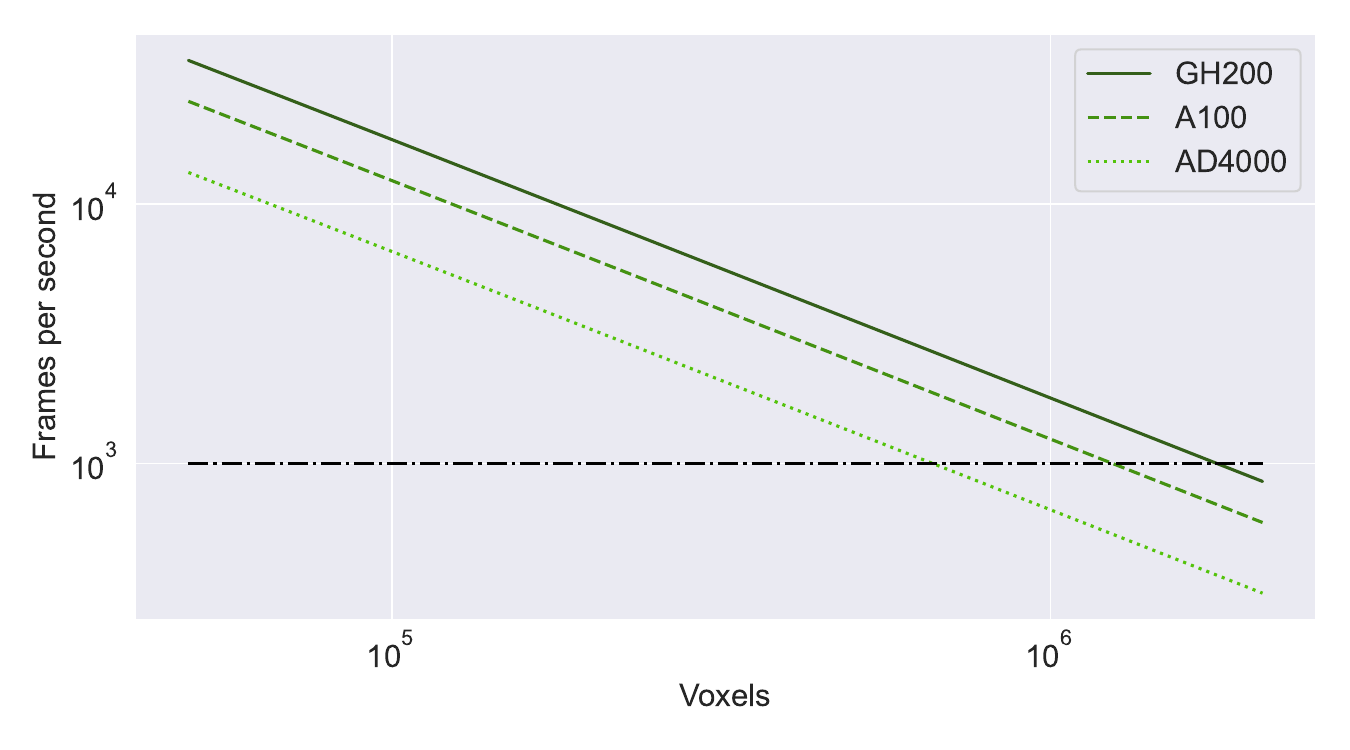}
\caption{Performance of beamforming for ultrasound. The number of voxels ranges from three orthogonal planes of $128\!\times\!128$ each, to the full $128^3$ data volume. The horizontal dash-dotted line indicates the minimum number of frames per second required for real-time performance.}
\label{fig:medical_realtime}
\end{figure}

In Fig.~\ref{fig:medical_realtime} we show the number of frames per second that the TCBF can sustain on different GPUs. The processing includes the 1-bit packing and transpose of the measurement matrix. It excludes these steps for the model matrix, as this typically happens once before the experiment and does not need to be repeated. For a set of three orthogonal planes, all three GPUs can easily sustain the required real-time frame rate of 1000 frames per second. None of the GPUs can process the full $128^3$ data volume in real time, although the GH200 is capable of processing $\sim85\%$ of the voxels in real time. Reducing for example the number of frequencies from 128 to 64 would make real-time processing of the full data volume possible for both the A100 and GH200.

\begin{figure}[tbp]
\includegraphics[width=\columnwidth]{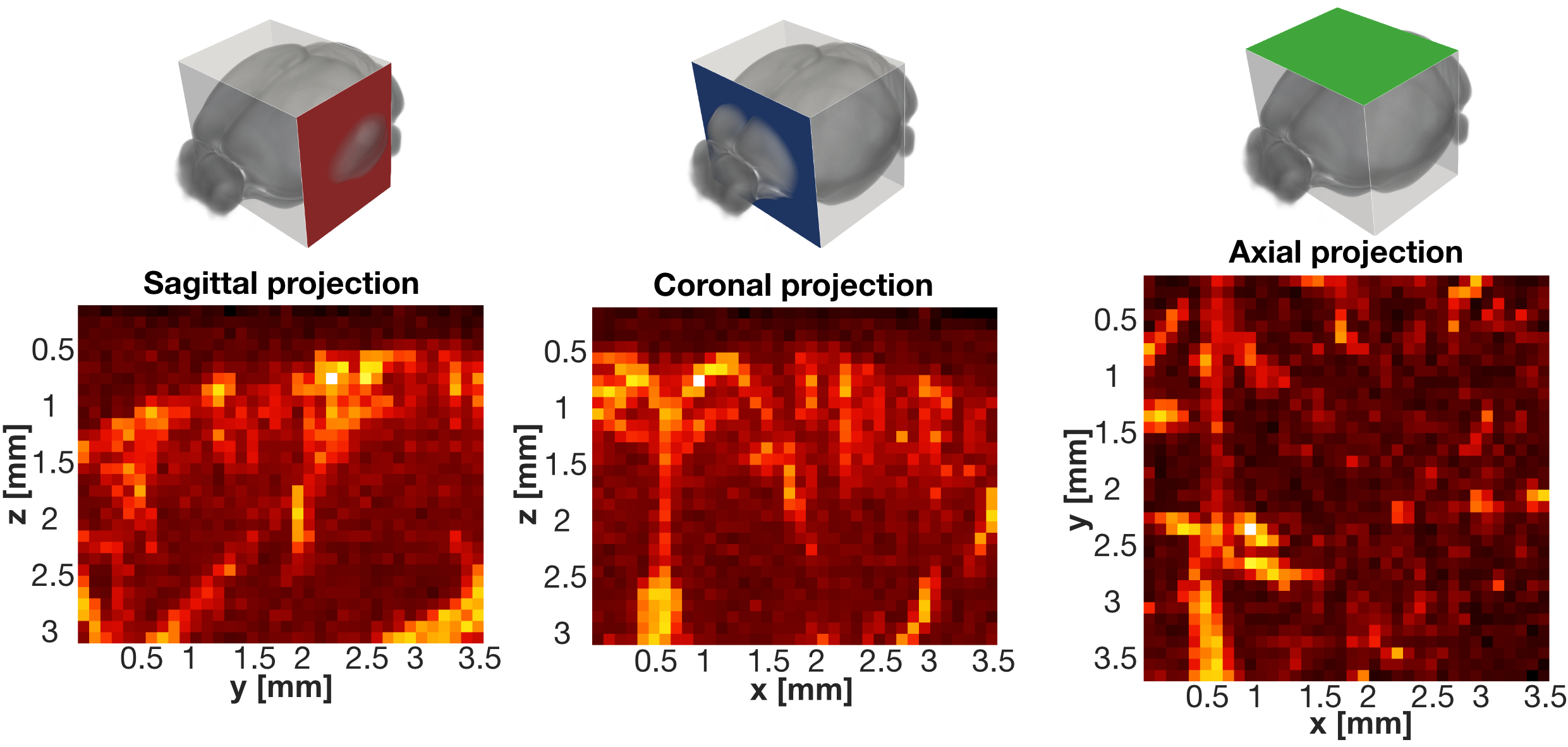}
\caption{Three orthogonal (sagittal, coronal and axial) maximum intensity projections through the beamformed volume. The volume which contains the blood flow inside a mouse brain was obtained by averaging the magnitude of the complex beamformed signal along the 8041 frames. See also \cite{BrownSciAdv2023}.}
\label{fig:medical_output}
\end{figure}

In addition to the real-time system, we explore the use of the TCBF for beamforming of pre-recorded data. In this case, there is no real-time constraint. However, quick feedback on experimental results is still important. As a dataset we use the anesthetized mouse brain dataset presented in~\citep{BrownSciAdv2023}. We beamform a subset of the volume, with a total of $36\times30\times30$ voxels. The dataset contains 8041 frames, each with 128 temporal frequencies, 64 transmissions, and 64 transceivers. This leads to a matrix-matrix multiplication with shape $M=38880$, $N=8041$, $K=524288$. Excluding reading the data from disk, the TCBF can process these data in $1.2\,$s, which significantly shorter than the real-time requirement of $8\,$s, leaving room for e.g. Doppler processing. Ultrasound processing is typically done in Matlab, Python or Octave. As a comparison, we run the matrix-matrix multiplication in float32 precision using Octave with OpenCL backend. On an A100, this takes roughly 15 minutes. The TCBF is nearly three orders of magnitude faster, allowing, for the first time, real-time feedback on such large volumes. While conversion to 1-bit means that the contrast is reduced, combining this much data still results in usable image feedback as shown in Fig.~\ref{fig:medical_output}.

\subsection{Radio astronomy}
In radio astronomy, beamforming techniques are employed to enhance signal detection by combining data from multiple antennas.

LOFAR (Low-Frequency Array)~\citep{vanHaarlem2013} is a radio telescope network consisting of tens of geographically distributed stations across Europe. Each station is composed of numerous individual antennas that collectively capture radio signals from the sky. These signals are initially processed by a station beamformer, implemented on Field-Programmable Gate Arrays (FPGAs) within each station. The station beamformer combines the signals from all antennas in the station into a coherent station beam, effectively pointing the station at a specific region of the sky. The resulting data, known as beamlet data, is then transmitted to a central beamformer~\citep{Broekema2018}, where the signals from all stations are coherently combined. This central processing step allows LOFAR to achieve high sensitivity and resolution, enabling detailed observations of astronomical phenomena across a wide field of view.

The central processing facility employs a second stage of beamforming, which can perform either coherent or incoherent beamforming. Coherent beamforming preserves phase information by aligning the signals from each station, producing a high-resolution, narrow beam with increased sensitivity. This approach is computationally intensive but is essential for high-angular-resolution observations, such as pulsar studies or imaging of compact objects. In contrast, incoherent beamforming discards phase information and instead combines the power from each station, creating a broader beam with a wider field of view but lower resolution. This method is computationally less demanding and is well-suited for all-sky surveys and transient detection.

When observing pulsars and fast transients with an interferometer, achieving high-time resolution is crucial. Typically, a time resolution of $\lesssim1\,$ms is used. To achieve such a high time resolution and retain manageable data rates, the spatial resolution is reduced.

This beamforming approach offers several advantages. Higher angular resolution enhances precise localization and background rejection. Additionally, each station's wide field of view enables high survey speeds, particularly when the entire field can be processed. Multi-beaming capabilities further enhance interference rejection and allow for the construction of a larger total collecting area. However, these benefits come with trade-offs, including a restricted field of view unless multiple beams are synthesized, potentially higher data rates due to numerous data streams, and the need for precise calibration to ``phase up" the array. Moreover, this technique can result in a complex instantaneous sidelobe pattern.

LOFAR beamforming is mapped to matrix-matrix multiplication as follows:
$M$ represents the number of beams, with each beam corresponding to a row in the resulting matrix. The parameter 
$N$ is the number of samples (in time), representing the number of columns in the output matrix. 
$K$ corresponds to the number of stations, reflecting the number of inputs combined during the matrix-matrix multiplication process. 
Finally, the product of the number of polarizations and channels is the batch size.

A LOFAR tensor-core beamformer is implemented using the 16-bit mode of \ccglib. As data are typically already GPU-resident and remain on the GPU for further computations, we only consider the matrix-matrix multiplication component for performance analysis. The chosen parameters are 1024 beams, 1024 samples, a range from 8 to 512 stations to be combined, and a batch size of 256. This configuration is also run using the reference LOFAR beamformer on an A100 GPU. It runs in float32 precision on the normal GPU cores. Note that we have removed the calculation of beamformer weights from the reference beamformer, to be able to fairly compare the reference and tensor-core implementations.

\begin{figure*}[tbp]
\includegraphics[width=\textwidth]{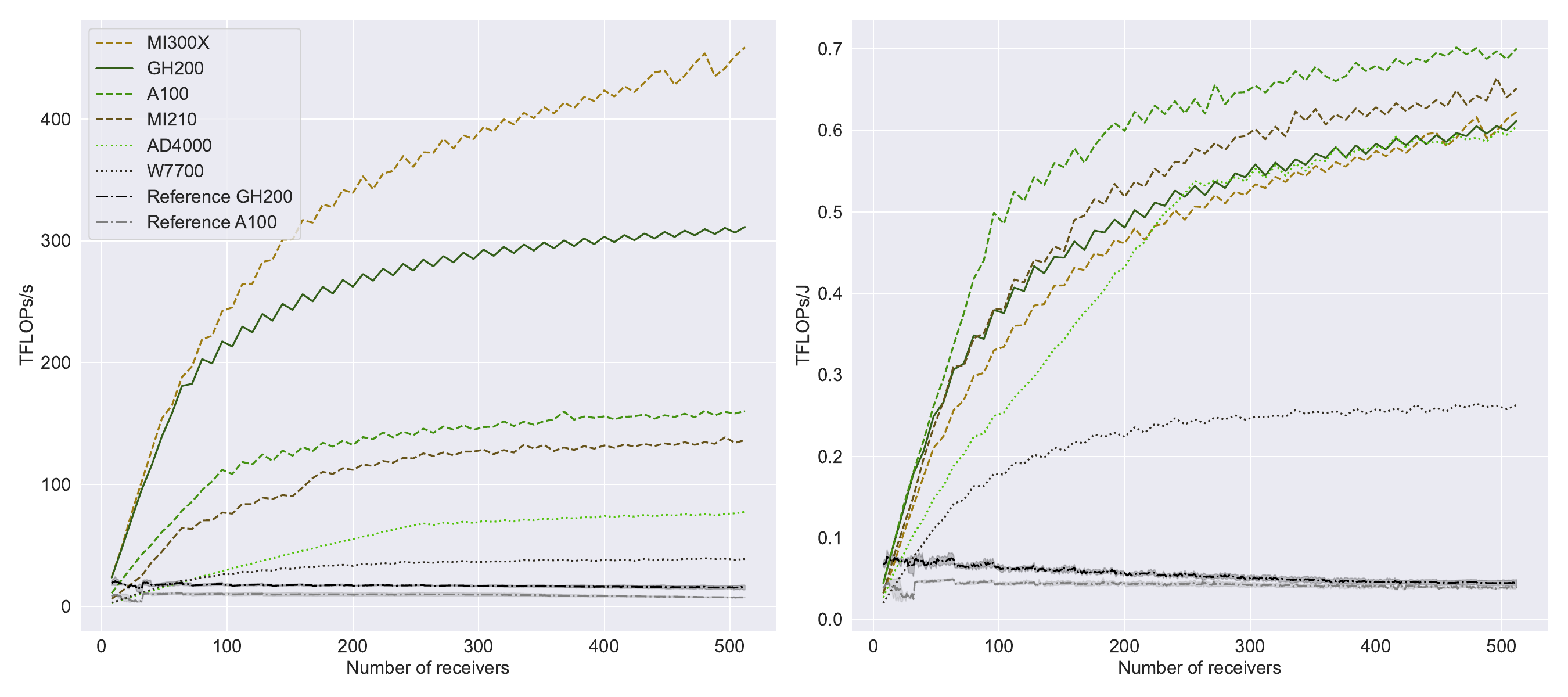}
\caption{Performance (left) and energy efficiency (right) of the LOFAR TCBF. The reference lines for A100 and GH200 correspond to the current LOFAR beamformer kernel (without Tensor Cores) running in float32 precision.}
\label{fig:lofar_bench}
\end{figure*}

The performance and energy efficiency of the LOFAR TCBF are shown in Fig.~\ref{fig:lofar_bench}. The sawtooth pattern stems from padding that happens when the number of receivers is not a multiple of the amount of work per GPU thread block set during the auto-tuning of the kernel. Except for very small numbers of receivers, the TCBF outperforms the reference beamformer on both the A100 and GH200, in both throughput and energy efficiency. On the A100, the TCBF is up to 20 times faster and 10 times more energy efficient than the reference beamformer. For the typical LOFAR configuration of 48 stations, the TCBF is still several times faster. The MI300X outperforms the GH200 on this application, achieving up to 50\% higher performance, with similar energy efficiency. However, with 512 receivers, the workload is still too small to fully saturate this GPU, preventing it from reaching its peak theoretical performance (approximately twice that of the GH200).

\section{Conclusions and future work}
\label{sec:conclusion}
We have introduced the Tensor-Core Beamformer, with at its core a high-performance and energy-efficient complex matrix-matrix multiplication library with outstanding performance on both NVIDIA and AMD GPUs. We have shown applications in both medical ultrasound and radio astronomy, where the 
TCBF improves significantly upon earlier beamformers in both performance and energy efficiency. In medical ultrasound, real-time imaging is essential, allowing a surgeon to change their course of action based on the ultrasound images. Because the TCBF is up to three orders of magnitude faster than previous implementations, this real-time feedback is now for the first time possible for 3D computational ultrasound imaging.
The radio-astronomical TCBF is 2-20 times faster than the existing beamformer, as well as 10 times more energy efficient. This makes it possible to either form more beams in real-time, or reduce the amount of hardware needed for beamforming, reducing energy consumption significantly as well.

Several improvements and extensions to \ccglib are being considered for future releases:
Firstly, the tensor cores support more precisions than just float16 and int1. Both NVIDIA and AMD (starting with CDNA3) support tensorfloat32, a 19-bit format with the same range as float32 but less precision. AMD supports float32 as well. Support for these formats is currently available as an experimental feature in \ccglib. The most recent architectures have introduced several 8-bit float formats, which may become relevant in the future.

Secondly, the matrix-matrix multiplication kernels in \ccglib currently require a transpose of the input data because the complex data have to be separated into their real and imaginary components, instead of the more usual interleaved storage format. In the future, we would like to provide a matrix-matrix multiplication kernel that does not require this transpose, and works on interleaved real and imaginary data instead. Such a method has already been used successfully in the tensor core correlator~\citep{Romein:21a}.

Lastly, for NVIDIA's Hopper and Blackwell generations, new interfaces were introduced for the tensor cores, along with enhancements like the tensor memory accelerator. To achieve maximum tensor core performance, these features must be leveraged. Supporting this in \ccglib is highly non-trivial, but will likely be important to maximize performance on future GPU generations.

\section*{Acknowledgments}
 This work was supported by the Netherlands eScience Center under grant number~ETEC.2020.025 (RECRUIT), by the RADIOBLOCKS grant (HORIZON-INFRA-2022-TECH-01, Grant Agreement nr. 101093934), and by the Netherlands Organization for Scientific Research (NWO) through the DAS-6 grant \cite{DAS6}. This work has made use of resources and expertise provided by SURF Experimental Technologies Platform (SURF-ETP), which is part of the SURF cooperative in the Netherlands, under project no. SURF-ETP0012.
 We would like to thank AMD for providing access to MI300 GPUs and NVIDIA for donating an A100 GPU.

\bibliographystyle{IEEEtran}
\bibliography{main}

\end{document}